\documentclass[10pt]{article}

\usepackage{amssymb}
\usepackage{graphicx}
\usepackage{authblk}
\usepackage{color}

\twocolumn

\begin{document}

\title{Measurement of electron-calcium ionization integral cross section using an ion trap with a low-energy, pulsed electron gun}

\author{{\L}ukasz K{\l}osowski \thanks{lklos@fizyka.umk.pl} }
\author{ Mariusz Piwi\'nski} 
\author{Szymon W\'ojtewicz}
\author{Daniel Lisak}
\affil{Institute of Physics, Faculty of Physics, Astronomy and Informatics, Nicolaus Copernicus University in Torun, Grudziadzka 5, 87-100 Torun, Poland}
\date{}

\maketitle

\begin{abstract}
An apparatus for production of various atomic and molecular ions inside a linear Paul trap has been set up. 
The system applies a custom-made, low energy, pulsed electron gun to produce ions in electron impact process. 
Such ionization method can find some interesting possible applications such as derivation of ions inaccessible other-ways, which can be used in molecular ion experiments. 
The technique allows also for determination of cross sections for various collisional processes. 

As a feasibility study, the apparatus was used for determination of ionization integral  cross section of calcium in the 16--160 eV range of electron impact energy. 
The obtained cross section  values are discussed and compared  with existing data sets.
\end{abstract}


\begin{figure*}[h!]
\includegraphics[width=\textwidth]{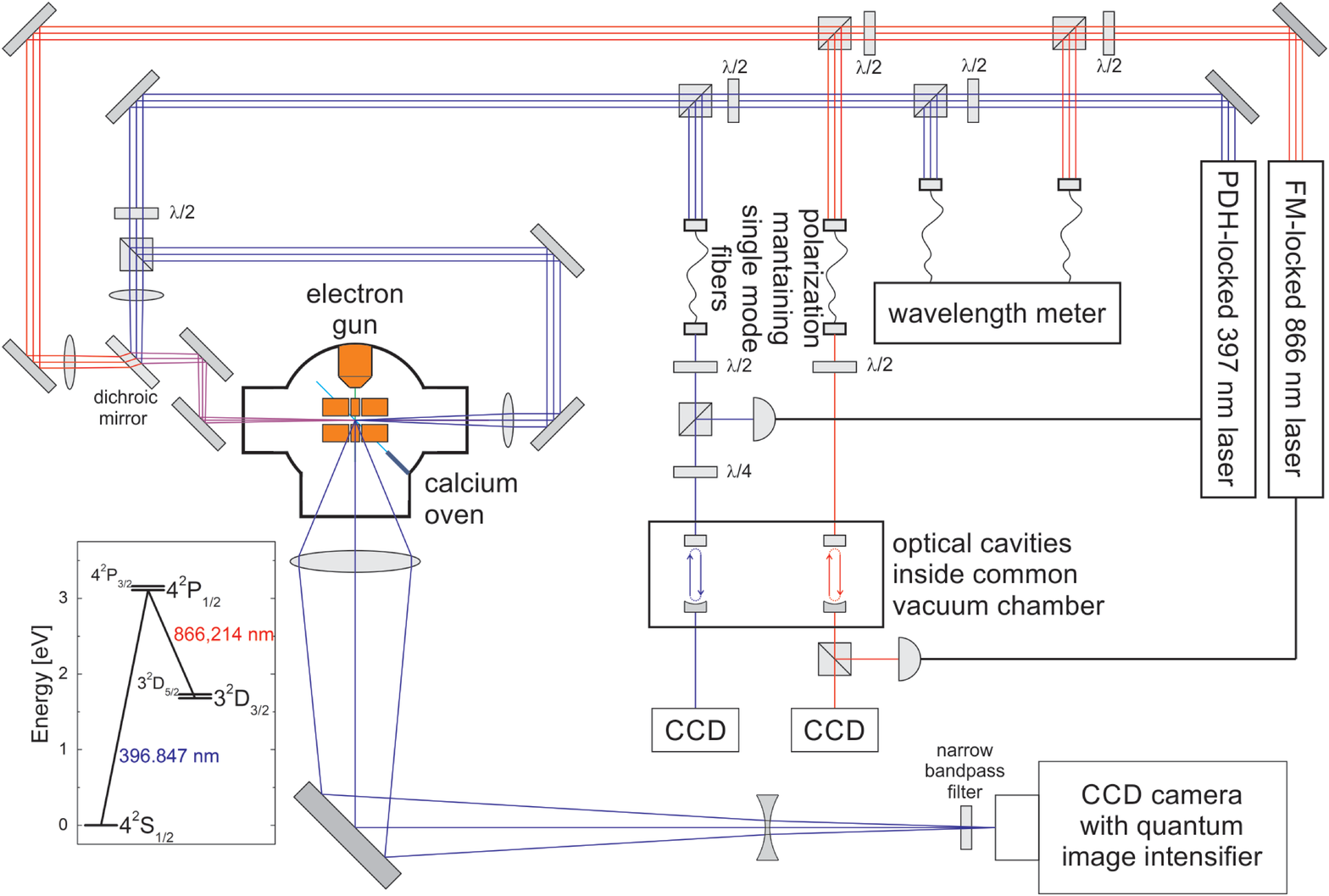}
\caption{Simplified scheme of the experimental setup. The laser beams are produced by Toptica DL pro (397 nm) and DL 100 (866 nm) diode lasers. The lasers are stabilized using high finesse optical cavities. 
Both beams are merged using dichroic mirror and introduced to the vacuum chamber and aligned to the trap's main axis. The beams are focused near the trap center. To balance the photon pressure, the 397 nm beam is splitted into two counter-propagating beams. 
The sources of atomic, molecular and electronic beams are aligned to cross the trap center. The trapped ions are detected using CCD camera equipped with two lenses, mirror, filter, and quantum image intensifier.  
The left bottom inset presents simplified scheme of the calcium ion energy levels with the optical transitions applied in the experiment.}
\label{schemat}
\end{figure*}

\section{Introduction}

Research involving trapped ions, at some stage of experimental procedure, requires loading the trap \cite{major}. 
It can be achieved by injection of ions from outside, which is possible using an external source \cite{otto13} or ablation of a solid with laser \cite{chen11,huds09} or electron \cite{troy15} pulse. 
An alternative way is creation of ions inside the trap by ionization of a neutral quantum object, such as atom or molecule \cite{kjae00,guld01,vers13,tong10}.
There are experiments, where some species of ions are derived from other ones in a chemical reaction \cite{wine87,molh00}.

The ionization utilizes usually a photon \cite{kjae00,guld01,vers13} or electron interaction with a neutral atom. 
Contemporary apparatuses involve rather photoionization than impact process as the laser technology experiences continuous development, allowing for driving optical transitions of an atom with high efficiency and precision, leading to well controlled production of ions. 
The most important disadvantage of such technique is that the optical ionization system can be designed for one, selected species of ions only. 
It is also difficult to apply in molecular ion production \cite{tong10}, as the optical spectra of molecules are much more complex than atomic ones.

More universal method of electron bombardment of neutrals  had been used in the past, 
however since the optical methods development it became much less common. 
As the electron impact ionization is not as selective as optical interactions, application of such technique can lead to ionization of a background gas, multiple ionization of atoms of interest, lack of isotope-selectivity of the process, etc. 
Additionally, most of the past applications of electron impact ionization inside trap involved relatively high energy of electron, of the order of several keV. 
The energies of such electrons are high above the ionization thresholds and typical electron impact ionization cross sections maxima of atoms and molecules. 
For example the maximum of cross section for calcium is noticed at around 25 eV \cite{cvej03}, typical molecules like CO$_2$ have their ionization maxima at about 100 eV \cite{huds04,stra96}. 
This way, the cross sections have usually low value at energies above 1~keV, making the electron impact inefficient for ion derivation.

In this paper we present a low energy (below 200 eV) electron gun system allowing for ionization of various neutral species at energies closer to ionization cross sections maxima. 
The selectivity of ion production can be achieved by application of proper trap settings \cite{werth}.
To improve the precision of electron beam controlling, a pulsed system has been designed. 
It provides better control of the number of produced ions and helps to avoid some undesirable effects such as heat transfer or interaction of electrons with ions of interest.

\section{Apparatus}
The apparatus has been set up in the National Laboratory FAMO in Torun. 
It consists of a vacuum chamber ($10^{-10}\div 10^{-9}$ mbar) equipped with a Paul trap \cite{paul90} and sources of necessary beams: electronic, atomic (calcium) and molecular (not used in the tests described in this paper). 
Additionally the system contains optical part used for cooling and detection of the trapped ions. 
Overall scheme of the apparatus is presented in figure \ref{schemat}.
The individual parts of the system are described in following sections of this paper.

\subsection{Linear Paul trap}
\begin{figure}[t]
\includegraphics[width=\columnwidth]{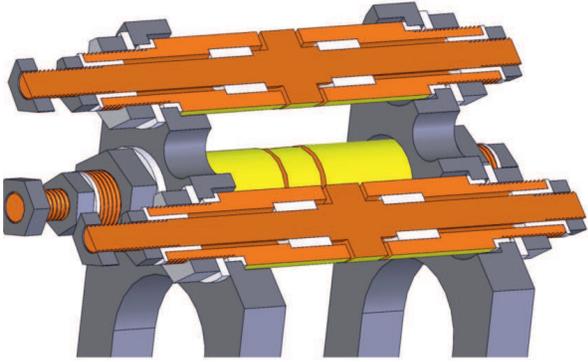}
\caption{Geometry of the trap used in the experiment (cross sections of 2 electrodes). The orange and yellow parts are the electrodes made of gold-coated copper. The gray parts are mechanical parts made of stainless, nonmagnetic steel. The white parts are insulators made of macor. }
\label{zdjecie}
\end{figure}

The  trap used in the system is a standard, linear, segmented Paul trap with cylindrical electrodes \cite{doug05}.  
Its geometry is presented in figure \ref{zdjecie}.
The cylindrical electrodes have diameter of 8.0 mm and are placed in a quadrupole configuration  with  4.0 mm distance  between the electrode pairs. To provide confinement of ions' motion along the trap axis, the electrodes have been divided into three segments. The central one has length of 4 mm, while the outer ones are 10 mm long. The gap between segments is 0.1 mm. 

The electrodes are held with conducting, grounded frame providing support for electric connections for the trap's voltage supplies and optical access for the laser cooling and imaging systems, as well as the atomic, molecular, and electronic beams.

The electrodes are made of copper cylinders (7.8~mm diameter) placed inside 0.1 mm thick gold tube (99.95\% purity), providing 8.0 mm diameter of the entire electrode. 
Such technique of coating allows to avoid flaking off the gold in vacuum, which is typical issue for layers obtained by vapor deposition.
The electric connectors are attached to the end part of the electrodes, outside the supporting frame. 
The central segment electrodes are connected coaxially with the outer segment parts.

As the intended way of ion production is electron impact, the trap's design requires that no dielectric surfaces are exposed to the stray electrons. 
Otherwise, there would be possible electric charge patches from electrons settling on such parts, disturbing the trapping potential, making proper operation of the trap impossible.
To provide such condition, all the insulating parts, made of macor, are hidden inside the electrodes, as presented in figure \ref{zdjecie}.

\begin{figure}[t]
\includegraphics[width=\columnwidth]{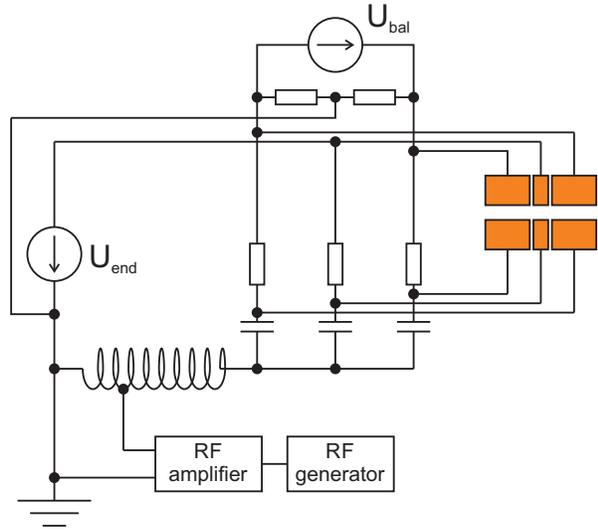}
\caption{Scheme of the voltage supply of the trap. Only 6 of 12 trap electrodes are shown. 
The other 6 are connected directly with the visible ones in a quadrupole configuration. 
Roles of the elements are explained in text. }
\label{zasilanie}
\end{figure}

The voltage supply system for the trap provides a radio-frequency voltage combined with DC voltages which are introduced to given electrodes via  electrical feedthrough. 
The schematic of voltage system is presented in figure \ref{zasilanie} together with notation of the voltages used. 

The radio frequency (RF) AC voltage is generated using a waveform generator (Agilent  33220A) and amplified with  high speed  amplifier (Falco Systems WMA-300) providing maximum output amplitude of 150 V. 
As at higher frequencies, higher amplitudes are necessary, additional amplification by helical resonator \cite{maca59} is introduced. 
Several, exchangeable resonators have been prepared to work with various frequencies in the range from 1 to 4 MHz with quality factor of the order of 10.

The DC voltages are provided by programmable power supplies (Tenma 72-2535). 
The $U_{end}$ ensures ion confinement along the main axis of the trap by forming of a minimum of static potential in the central part.
The $U_{bal}$ is used to control the axial position of the trapping potential minimum. 
Due to possible imperfections in trap's construction, such balancing can protect ion ensembles from being lost in the case of low trapping potentials.

At some electrodes, superposition of DC and radio frequency AC voltages is required, which is achieved by the use of independent $RC$ mixers. 

Typical trapping conditions were $1\div 20$ V of $U_{end}$, $0\div 100$ mV of $U_{bal}$, $100\div 500$ V of AC peak-to-peak amplitude at frequencies from 1 to 4 MHz (monitored continuously with an oscilloscope).

\subsection{Optical system}
The optical system \cite{klos17a} consists of two main parts: the Doppler cooling lasers and imaging system. The scheme of the optical system is presented in figure~\ref{schemat}.

\subsubsection{Cooling of calcium ions} 
Cooling is performed in a well known Doppler scheme \cite{hans75} using two laser beams. 
The simplified graph of energy levels of Ca$^+$ ion is presented in figure \ref{schemat}.
The Doppler cooling involves \mbox{$4^2S_{1/2}\leftrightarrow 4^2P_{1/2}$} transition at wavelength 396.84673~nm \cite{nist}. 
To avoid trapping of the ions in $3^2D_{3/2}$ dark state, it is necessary to couple optically the $4^2P_{1/2}$ and $3^2D_{3/2}$ states at wavelength 866.2140~nm \cite{nist}.
In our case, lasers are Toptica DL pro and DL 100, stabilized using external optical cavities in a Pound-Drever-Hall \cite{drev83} and FM-lock \cite{smit70} scheme respectively.
Both optical beams are merged together with a dichroic mirror and aligned in the main axis of the trap with the focal points in its center.
Spot sizes of the beams in this point are approximately 500 $\mu$m with typical powers of several milliwatts.

Additionally a weak magnetic field (0.5 mT) was used for mixing the $D$ sub-states to avoid ion imprisonment in the dark state.
The source of the magnetic field is a set of permanent neodymium magnets arranged in a proper geometrical configuration.
The magnetic field is oriented along the axis of the electron gun to provide optimum performance of electronic beams. 

In the case, when some other species of ions are trapped besides calcium, their slowing down will be based on a well known effect of sympathetic cooling \cite{rohd00}. 
Briefly, if there are various ions trapped, they interact by a Coulomb force continuously exchanging momenta and energies. 
Thus, if just some subset of ions is directly cooled in optical way, the others will also loose their kinetic energy in such exchanges.
As the sympathetic cooling is effective for interaction between objects of similar mass-to-charge ratios, 
Ca$^+$ ions (40 a.u.) can be used to cool large variety of molecular ions such as CO$_2^+$, CO$^+$, N$_2^+$, O$_2^+$ 
 or various organic molecules. 

\subsubsection{Imaging system}
Imaging is based on the detection of 397 nm fluorescence from the Doppler-cooled Ca$^+$ ions.
The camera optics is aligned perpendicular to the main axis of the trap and focused on its central part.
The images are captured with  camera (PCO. Dicam Pro) equipped with a quantum image intensifier and a narrow bandpass filter (395 nm central wavelength, $\pm10$ nm bandwidth). 
The focusing optics consist of two lenses as presented in figure \ref{schemat} and a mirror allowing improved alignment of the imaging path.
The system provides resolution of approximately 2~$\mu$m/pixel.

\subsection{Source of atomic beam}
Production of ions in the presented apparatus requires ionization of a neutral object such as molecule or atom.
Such atoms can be introduced in the trap by forming a beam of appropriate geometry and density.

The  calcium beam is produced with a custom made oven. 
Its main part is a steel tube, 80 mm long and 2~mm in diameter, 
filled with calcium powder in approx. 1/3 of its length. 
The calcium-free part is used to form an effusive beam.
Half of the tube's length  is helically covered with heating wire (Thermocoax 1HcI15, 12.4 $\Omega$/m,  1.5 mm  of outer diameter) providing approximately 3.5 $\Omega$ of electric resistance. 
The working temperature of the oven is set from 200 to 500 $^\circ$C controlled by DC current ($0.4\div 1$~A).
The oven is placed in a steel shielding to reduce its thermal radiation.

\section{Pulsed electron gun}

\begin{figure}[h!]
\includegraphics[width=\columnwidth]{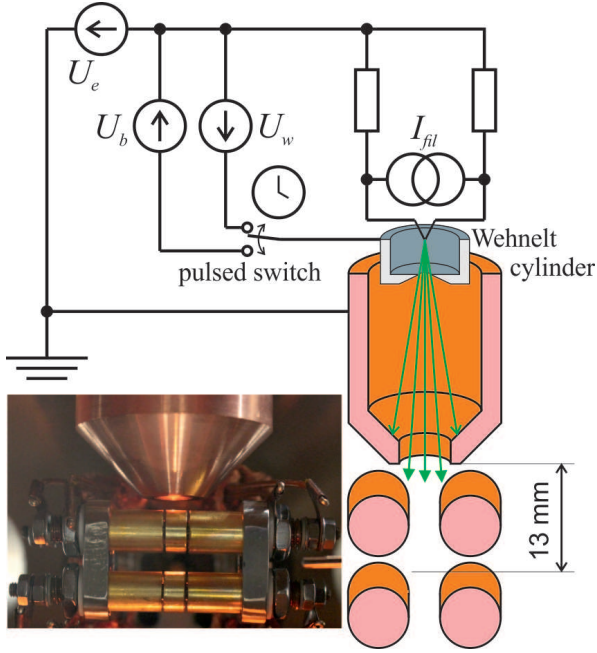}
\caption{Schematic of the pulsed electron gun and its voltage supply system. The energy of electrons is defined with $U_{e}$ voltage. 
The transmission of electrons through Wehnelt cylinder is enabled/disabled by switching its potential between $U_{w}$ (working Wehnelt potential) and $U_{b}$ (blocking barrier potential).  
A time relay, switching  between the two voltages, is applied to control the pulse duration. 
The filament, being the primary source of electrons, is heated up with electric current from $I_{fil}$ supply.
Photograph of the gun with the trap is added to view geometrical relations between the elements.}
\label{elektryka}
\end{figure}

Custom made pulsed electron gun was used as a source of electronic beam. 
Its schematic diagram with voltage supply system is presented in figure~\ref{elektryka}.

The primary source of electrons is thermal emission from a tungsten  filament (Kimball Physics Inc. ES-020) heated up with current of 2.3 A. 
It is housed in a Wehnelt cylinder and followed by an 80~mm long, cone shaped, grounded shield with aperture of 5 mm in diameter, providing efficient screening from the trap's electric field. 
Thus, the kinetic energy of electrons leaving the gun is defined by potential of the filament's tip.
The beam is geometrically formed by aperture of the shield, which is localized 13 mm above the center of the trap.

A solution applied in the electron gun is a programmable pulse system allowing for creation of various series of electron pulses, also single ones. 
As the number of ions captured in the trap depends on time of duration of the electron pulse and there are numerous effects resulting from electron-ion interactions, this is necessary to well control the pulse time.

The gun should be able to be switched ON/OFF in a short time. This way, heating up and cooling down the filament cannot be applied due to thermal inertia of gun elements. 
Switching the gun can be however easily realized by switching potentials of the Wehnelt cylinder between two states:
\begin{enumerate}
\item {ON} state, when the Wehnelt cylinder is kept on potential $U_{w}$, several volts above the potential of the filament's tip. Optimum setting can be found experimentally by optimizing electron current recorded on the trap's electrodes,
\item {OFF} state, when the Wehnelt cylinder has low potential $U_{b}$, several tens of volts below the filament's tip, forming a barrier for the electrons. 
\end{enumerate} 
The switching is realized with a commercial  time relay (COBI ELECTRONIC, CRT-V1).
It is programmed to form a single logic pulse triggered by a manually operated switch. 
The device allows for pulse duration from 0.01~s to 100 hours with 10 ms accuracy. 
The typical values for the experiment are in the range from 0.1 to 1.0 s.

\subsection{Tests of the electron source}
To calibrate the efficiency of the gun, some questions need to be addressed: total efficiency of the gun versus electron energy, probability of passing the trap center by electron and the gun's time characteristics.

As charged particles yield the trap's field, one should be aware of possible behavior of particular electrons forming the beam when passing the trap's area:
\begin{itemize}
\item an electron can be repelled by the trap's field and hit the vacuum chamber wall or other element,
\item an electron can settle on trap's electrode -- such electrons can be easily detected by measurement of electric current from the electrodes,
\item an electron can pass through the trap without hitting electrodes and without passing the ion trapping region, terminating its path on some other, grounded element of the vacuum system,
\item an electron can pass through the trapping region with or without subsequent hitting the electrode -- such electrons are the ''useful'' ones, as they can ionize target atoms or molecules.
\end{itemize} 
The numbers of electrons in these groups depend on electron energy, electronic beam current and geometry, trapping conditions (RF amplitude and frequency, DC potential) and magnetic field present inside the vacuum chamber. 
The beam's current is a function of electron energy, depending on the gun's design.
It is in general difficult to predict theoretically, however it can be found experimentally.

To determine the gun's emitted electron current, two tests were performed.
Electric current $I_{coll}(E)$ from electrons collected by the trap's electrodes with the voltage supplies turned off, was measured with digital multimeter (Agilent 3458A). 
The measurements were performed for various electron energies from 15 to 105 eV and for two values of magnetic field, estimated from magnetic moments and positions of permanent magnets used as a source.
The measurements were conducted using long pulses (several minutes) for beam stability reasons.

\begin{figure}[t]
\includegraphics[width=\columnwidth]{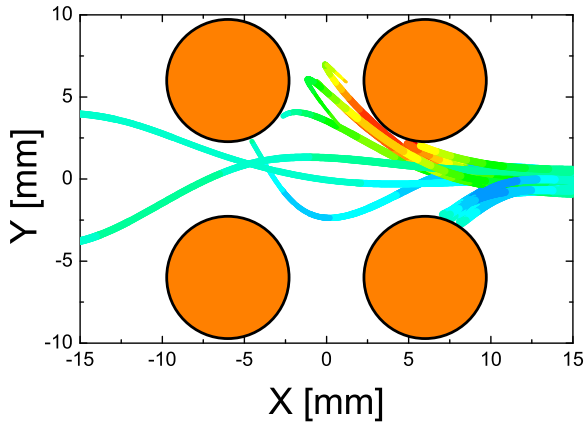}
\caption{ Example 20 simulated trajectories of electrons passing the trap (cross section perpendicular to the trap axis). The electron beam source is on the right of the picture. The settings were: the electron initial energy 70 eV, the magnetic field 0.5 mT, AC with frequency of 2 MHz and peak-to-peak amplitude of 150 V, the $U_{end}$ value is 15 V.
Thickness of the trajectory describes distance from the symmetry plane of the trap in the direction perpendicular to the plane of the picture. The kinetic energy of electron is described with color from red (200 eV) to blue (0 eV).}
\label{example}
\end{figure}

The $I_{coll}(E)$ readouts were followed by Monte-Carlo simulations of electrons' trajectories using custom-made Fortran procedures. 
The random parameters were the initial state of an electron (at selected energy, positioned in the gun's exit aperture).
As a result one obtains a large set of electron trajectories, similar to ones presented in figure \ref{example}.
The fraction $\eta$ of electrons that in simulations can hit an electrode is the searched parameter. 
The total electronic current can be then estimated from:
\begin{equation}
I_{tot}(E)=\frac{I_{coll}(E)}{\eta(E)}.
\end{equation}
\begin{figure}[t]
\includegraphics[width=\columnwidth]{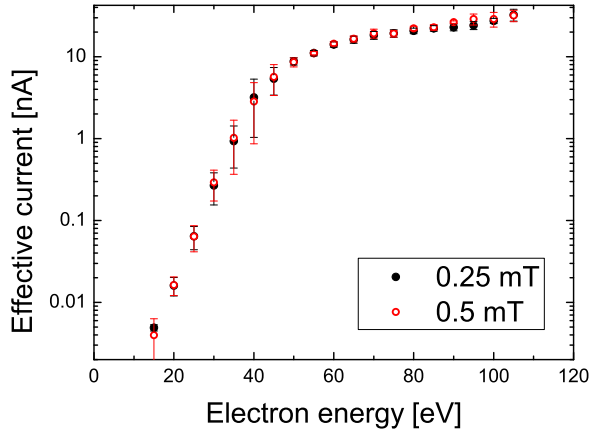}
\caption{ The effective electron current passing the trap versus the electron energy. The data were obtained by measuring the current collected by trap's electrodes and performing numerical simulations (see text for details). The measurements were performed for two various magnetic field values.}
\label{montecarlo2}
\end{figure}
\begin{figure}[h!]
\includegraphics[width=\columnwidth]{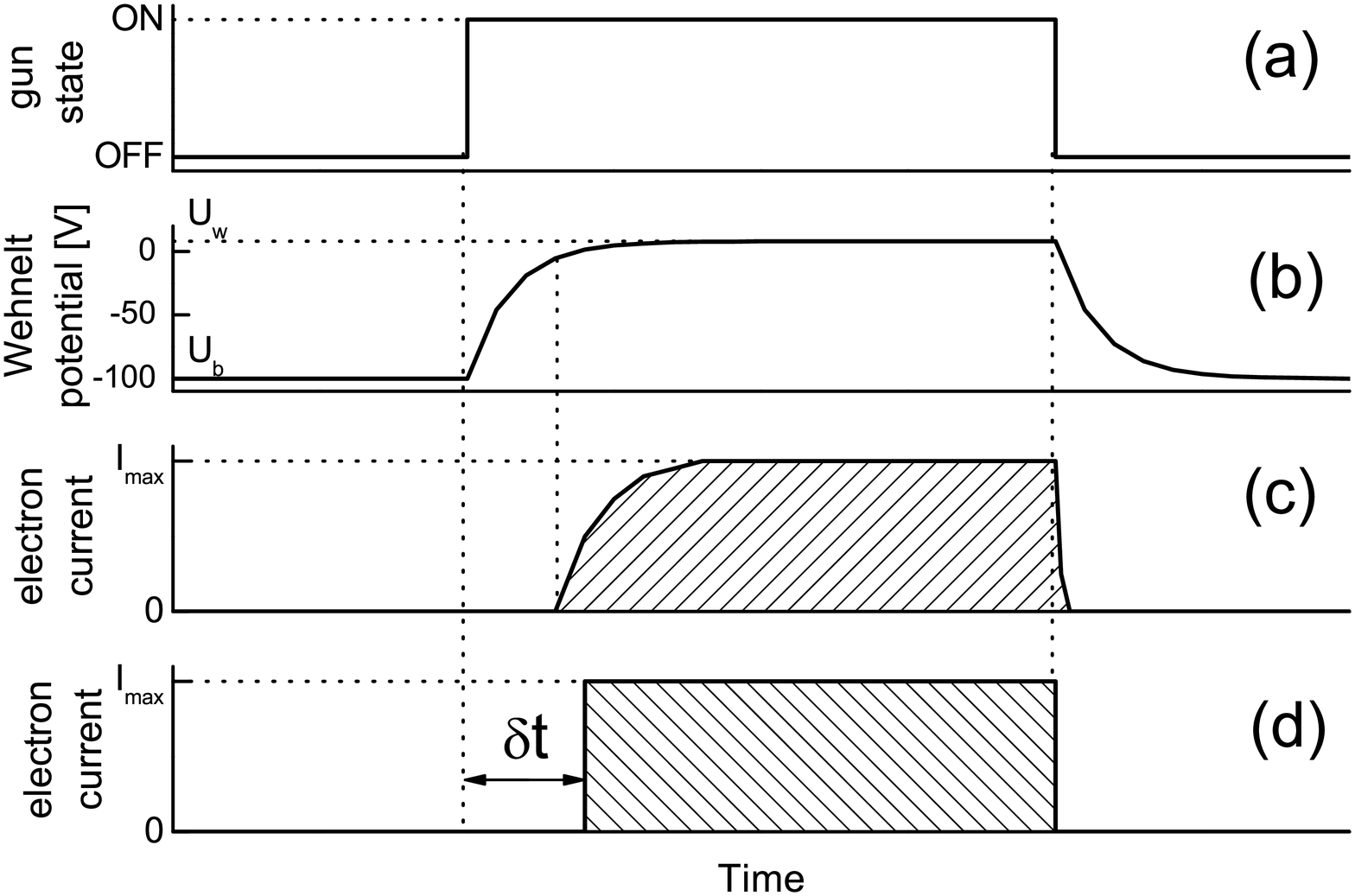}
\caption{Time characteristics of the electron gun. 
The (a) panel is gun's state switching between $U_b$ and $U_w$ voltage supplies. 
The (b) panel represents the response of Wehnelt cylinder including capacity effect of charging/discharging of the electrode in a transient state. 
The (c) panel represents the expected electron beam intensity. 
The (d) is a perfect rectangular pulse transmitting the same amount of charge as the real pulse (hatched areas are the same in both (c) and (d) panels).
The dead time is difference between pulse durations in panels (a) and (d).}
\label{wykresy}
\end{figure}
As the total electron current is determined, the Monte-Carlo simulations can be repeated with the trap's field turned on. 
It is assumed that turning the trap ON or OFF has no influence on the performance of electrons before passing the gun's exit aperture.
The additional random parameter in such calculations is the phase of AC field, as the electrons are continuously emitted by the gun.

The searched parameter is a fraction $\kappa$ of electrons passing the central part of the trap, where ions can be created and trapped. 
For the calculations, the electrons are classified as ''useful'' in the case they cross a sphere of 2 mm in diameter placed in the trap center. 
Knowing total number of electrons, one can estimate the number of useful ones and determine the effective current:
\begin{equation}
I_{eff}(E)=\kappa(E) I_{tot}(E)=I_{coll}(E)\frac{\kappa(E)}{\eta(E)}.
\end{equation}
Typical values of $\kappa$ and $\eta$ are of the order of 1\% and 50\%, respectively and their values depend on the electron energy and settings of the trap.
Exemplary results of efficiency measured this way are presented in figure \ref{montecarlo2} ($\Omega=2.11\times2\pi$ s$^{-1}$, 200 V of RF amplitude, $U_{end}=15$ V). 
The conclusion from the image is that magnetic field, in the range used for experiment, does not influence significantly the effective current. 
Moreover, the electronic beams at energies below 15 eV can be too weak for further applications.

\begin{figure*}[t]
\includegraphics[width=0.24\textwidth]{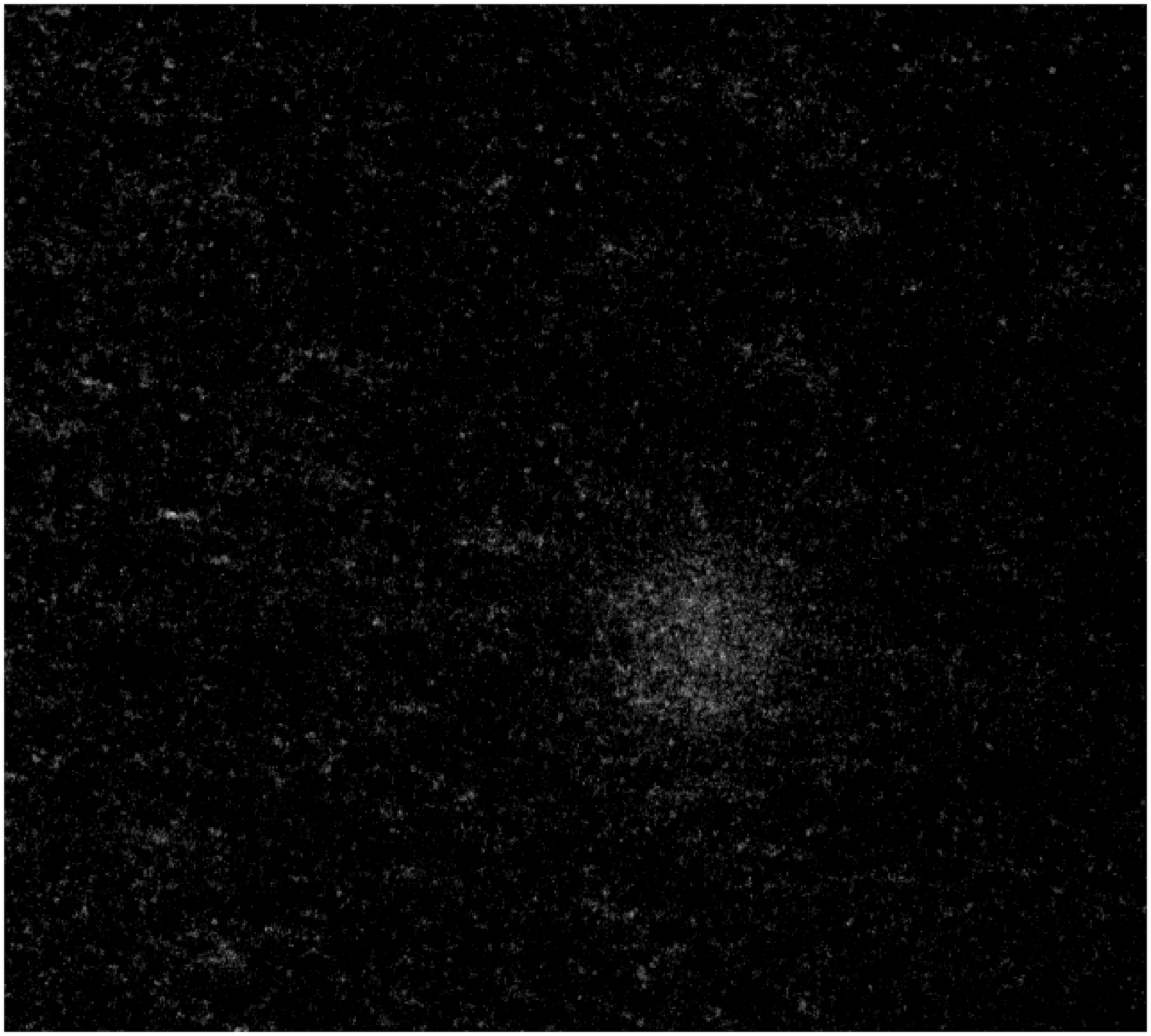}
\includegraphics[width=0.24\textwidth]{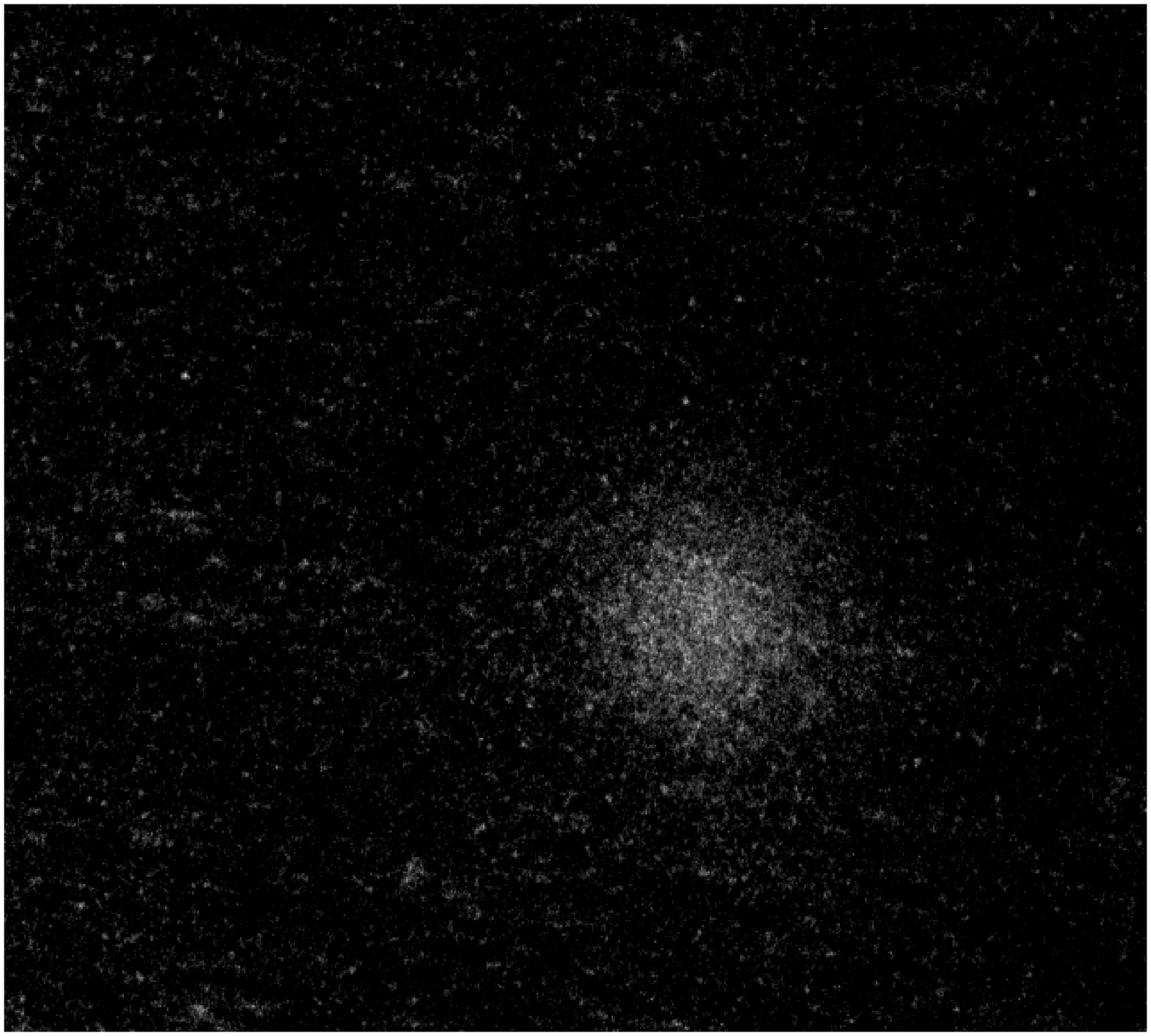}
\includegraphics[width=0.24\textwidth]{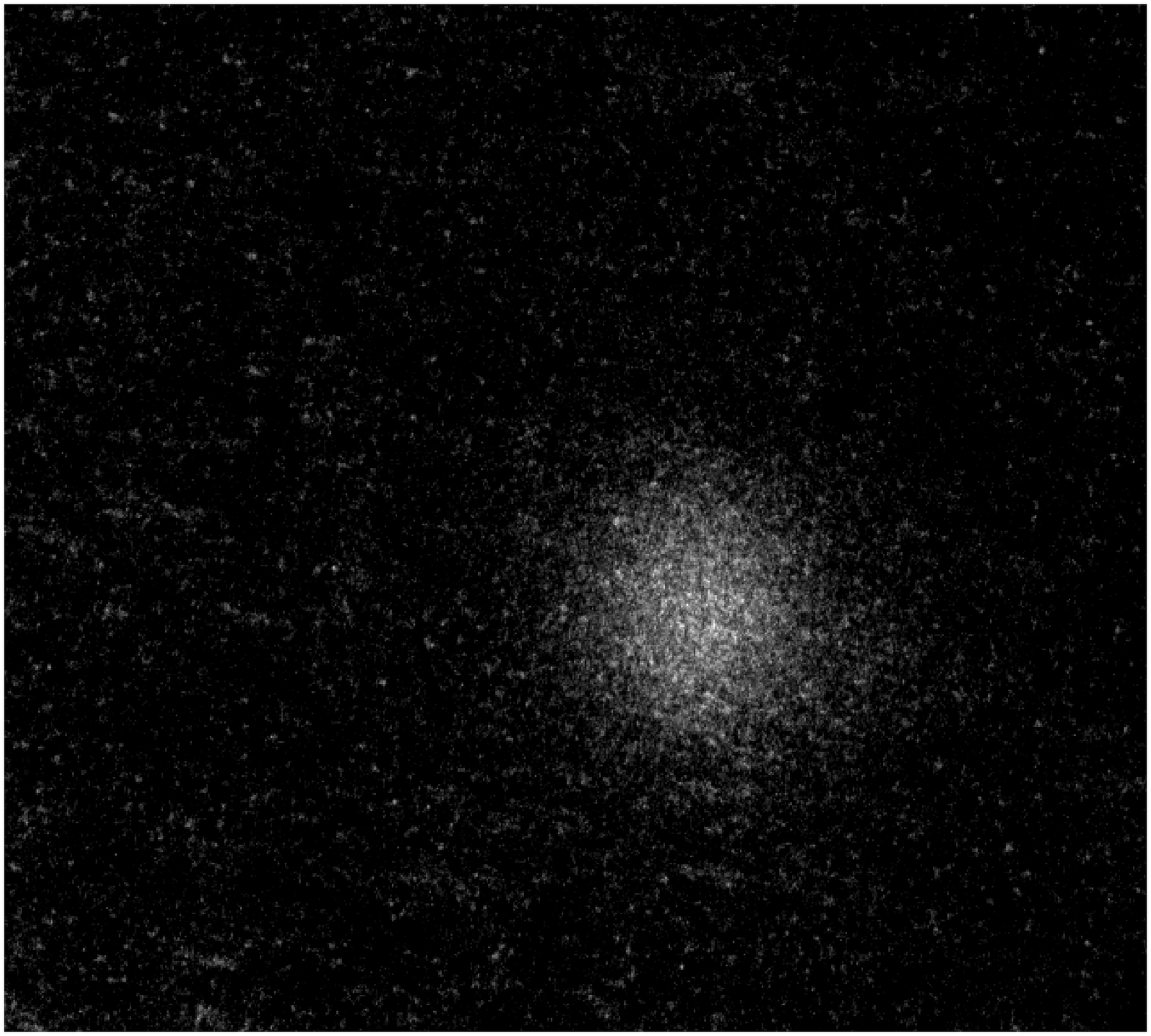}
\includegraphics[width=0.24\textwidth]{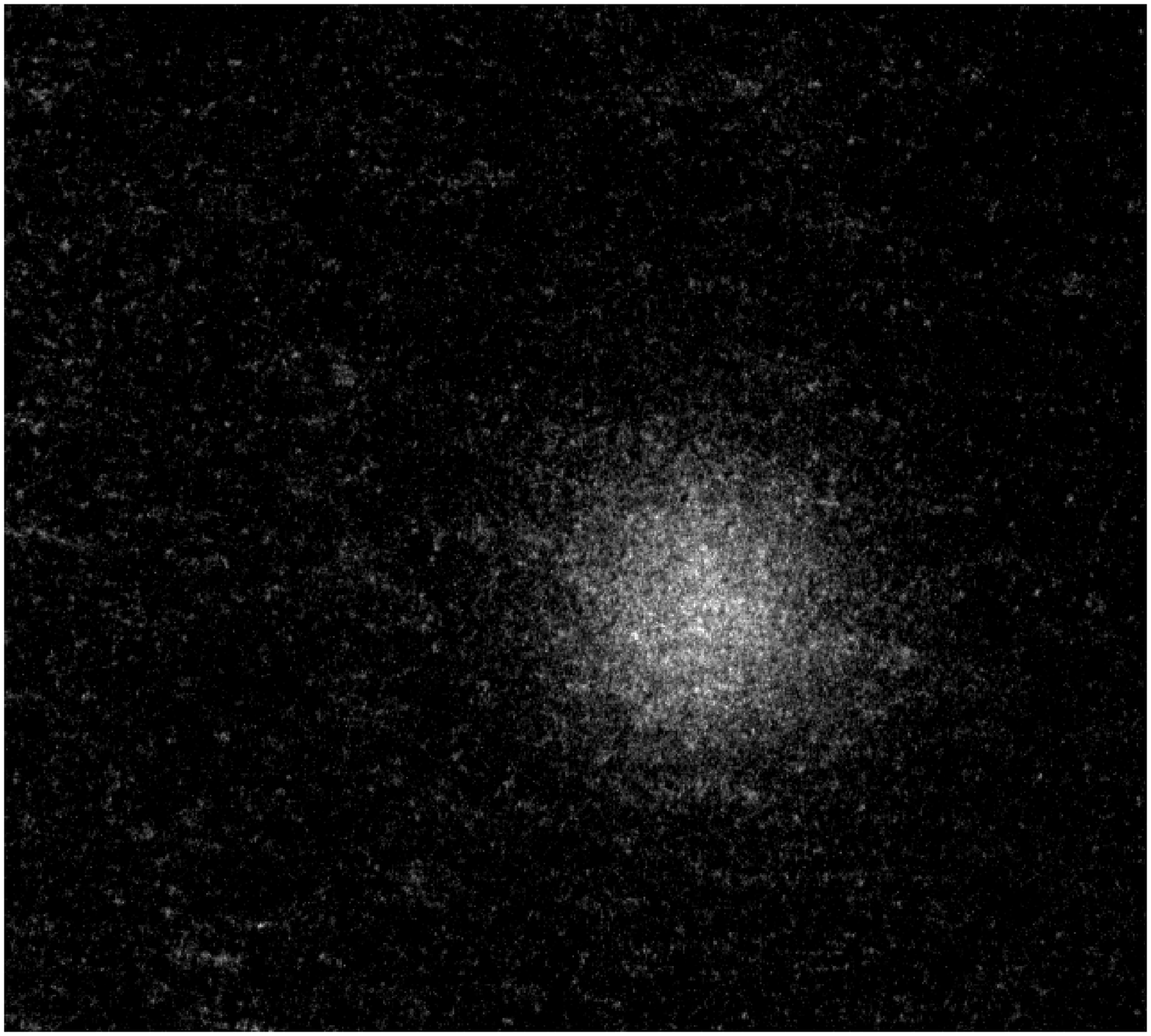}
\caption{Example images of ion clouds obtained for various electron pulse duration: 0.1 s, 0.2 s, 0.3 s and 0.4 s. Other parameters are explained in text.}
\label{zdjecia}
\end{figure*}
Another parameter is the time characteristics of electronic beam. 
Briefly, the electron gun remains in some transient state immediately after switching it ON or OFF, mainly due to finite electric capacity of the gun's electrodes. 
This way, the electron current is affected by such state and the effective pulse length is shorter than expected by some constant amount of time, which can be called a dead time and denoted with $\delta t$. 
Example time characteristics of the electronic pulse is presented in figure \ref{wykresy}. 
The electrons are transmitted through the Wehnelt cylinder only at sufficiently high potentials with optimum transmission at $U_w$.

To determine the dead time value, the ions were trapped in stable conditions of calcium oven, trap potentials, and electron energy. 
The trapping procedure was repeated for various pulse durations. 
The example images of ion clouds trapped in various conditions are presented in figure \ref{zdjecia}.

\begin{figure}[h]
\includegraphics[width=\columnwidth]{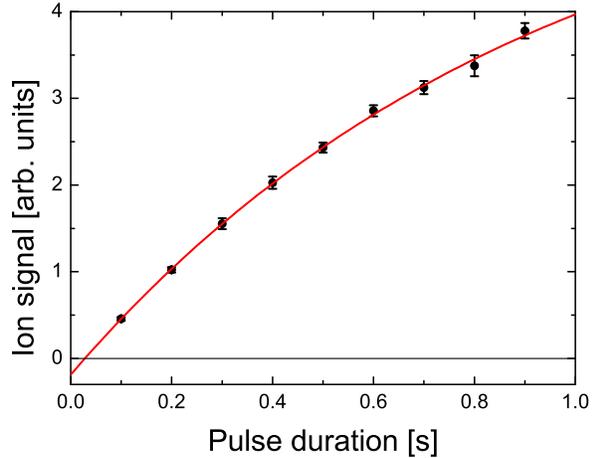}
\caption{ Image total intensity (proportional to number of ions captured) plotted versus the pulse duration (example for chosen settings of trap and electron gun). The data have been fitted with the function from eq. (\ref{4}) (solid line).  }
\label{pulse}
\end{figure}

The experiment was performed several times at various trapping conditions providing qualitatively similar results. 
Thus, only one of the data sets is analyzed here in details. 
The ions were trapped at AC frequency of 2.11 MHz and 200 V  amplitude and the $U_{end}$ of 15 V. 
The electron energy was 106~eV and the calcium oven was heated up with 800 mA current providing atomic density estimated to be approximately $10^8$ cm$^{-3}$.
The ions were Doppler cooled to temperatures of the order of 1 mK without Coulomb crystallization, which would be problematic with large ion ensembles.
The numbers of trapped ions were estimated from the intensity of images captured by the camera.
An example dataset presenting dependence of the ion number from the pulse length is shown in figure~\ref{pulse}.

For short pulses, where electron-ion interactions are negligible, the number of ions trapped can be expressed using equation:
\begin{equation}
N(t)=\frac{I_{eff}}el\sigma  n  (t-\delta t) ,
\label{3}
\end{equation}
where
$e$ is the elementary charge,
$\sigma$ is integral cross section for electron impact ionization of atoms at selected energy, 
$l$ is effective length of scattering region,
$n$ is number density of the atoms inside trap, 
$t$ is pulse duration, and
$\delta t$ is the dead time.
For longer pulses, when electron-ion interactions can lead to loss of some ions from the trap, the equation \ref{3} should be further developed to the form of:
\begin{equation}
N(t)=\frac{I_{eff}}el\sigma  n \tau  \left( 1- \exp\left(\frac{\delta t-t}\tau \right)\right) ,
\label{4}
\end{equation}
where $\tau$ is the time of saturation of ion number at certain experimental conditions.
The $\left(\frac{I_{eff}nl}e\sigma\right)$, $\tau$ and $\delta t$ parameters can be determined using least square fitting of the data from figure \ref{pulse} with equation (\ref{4}).

The obtained parameters can be used to estimate electron impact ionization integral cross section $\sigma$. 
As the efficiency of optical imaging system cannot be determined exactly as well as exact number density of atomic beam, the $\sigma$ results are obtained in arbitrary units and their values should be normalized using existing data sets for cross sections.

\begin{figure}[t]
\includegraphics[width=\columnwidth]{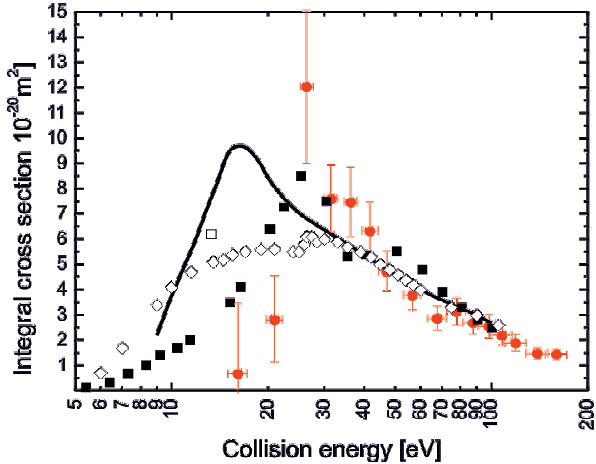}
\caption{ Integral cross section for electron impact ionization of calcium atom. (\textcolor{red}{$\bullet$}) -- present results, ($\diamond$) -- Okudaira \cite{okud70}, ($\square$) -- Okuno \cite{okun71}, ({$\blacksquare$}) -- Cvejanovi\'c and Murray \cite{cvej03}, (---) -- Fiquet-Fayard and Lahmani \cite{fiqu62}.}
\label{ics}
\end{figure}

\section{Calcium ionization cross section}
To test the ability of the apparatus to measure the electron impact ionization cross sections, some preliminary measurements for calcium were performed. 
Applying the procedure described in previous section and using values of effective electron currents, integral ionization cross sections were found for several electron energies from 16 to 160 eV. 

As only relative values of cross sections were measured,  the results were normalized to the experimental 100.5 eV point of Cvejanovi\'c and Murray \cite{cvej03} (the lowest uncertainty data points among energy-overlapping ranges of both experiments). 
The results are presented in figure \ref{ics}.

The ICS error bars are relatively large at lower energies, which is mainly due to low values of electron current. 
The energy error bars describe the energy spread introduced by trapping field and their size increases with the energy reaching over 10 eV at highest energies. 
Such uncertainty can be reduced using pulsed trapping field \cite{kjae03}, which requires  application of specially designed voltage supply system.
Such development of the apparatus is planned for our future research. 

The  relatively large energy uncertainty does not affect ICS results significantly in this range, as there are no pronounced structures in cross section functions for larger energies.

The obtained ICS results are in general in good qualitative agreement with existing data sets \cite{cvej03,okud70,fiqu62}. 
The energy of ICS maximum reproduces data of Cvejanovi\'c and Murray \cite{cvej03} and is quite close (slightly higher energy) to the maxima of Okudaira et al. \cite{okud70} and Fiquet-Fayard and Lahmani \cite{fiqu62}. 

The monotonic descent of cross section function at larger energies is consistent with all available data sets.

As the measurements at lower energies, around the maximum and below it, suffer from relatively large uncertainties, they should not be considered reliable reference for absolute values of integral cross sections.

\section{Summary and conclusions}
The presented experimental set up enables relatively easy ionization of various species of neutral atoms and molecules for applications in a linear Paul trap.
The system involves direct ionization of neutral atoms or molecules using an electron gun at low energies, close to the maxima of relevant ionization cross sections. 

Besides being an effective way of ion production, the apparatus allows to estimate values of integral ionization cross sections for the atoms or molecules used in further ion experiment. 
The cross sections for calcium were found as an example. 
The other species would require more complex method for their identification, which can involve for example  Coulomb crystallization of multi-species ion ensembles containing calcium. 

The most important advantage of the proposed method is its ability of production of wide choice of ions by changing only the gas used as a target for ionization. 
In case of photoionization, it would be very challenging due to complex energetic structure of molecules and required adjustments of laser systems.

The main issue of impact ionization is non-selectivity of such process, which can possibly lead to presence of unintended ion species in the trap. 
Such disadvantage can be however overcome by careful tuning of the trapping parameters allowing for mass selection of the trapped ions \cite{werth} using for example nonlinear resonance \cite{klos18} to eject ions of chosen masses.

\section*{Acknowledgements}

This work has been supported by the National Science Centre, Poland,  project no.~2014/13/B/ST2/02684.

\bibliographystyle{nature} 

\end{document}